\documentclass[10pt, conference]{IEEEtran}

\usepackage{flushend}
\usepackage{float}

\usepackage[shortlabels]{enumitem}

\usepackage{booktabs}
\ifCLASSOPTIONcompsoc
  \usepackage[nocompress]{cite}
\else
  \usepackage{cite}
\fi
\ifCLASSINFOpdf
   \usepackage[pdftex]{graphicx}
  \DeclareGraphicsExtensions{.pdf,.jpeg,.png}
\else
\fi
\usepackage[cmex10]{amsmath}
\usepackage{algorithmic}
\ifCLASSOPTIONcompsoc
  \usepackage[caption=false,font=footnotesize,labelfont=sf,textfont=sf]{subfig}
\else
  \usepackage[caption=false,font=footnotesize]{subfig}
\fi
\usepackage{stfloats}
\usepackage{url}


\begin{document}
%
\title{A Scalable Model for \\Secure Multiparty Authentication}

\author{\IEEEauthorblockN{Hussain Al-Aqrabi and Richard Hill}
\IEEEauthorblockA{Centre for Industrial Analytics\\
University of Huddersfield\\
Huddersfield, HD1 3DH, UK\\
Email: \{h.al-aqrabi, r.hill\}@hud.ac.uk}
}


%


\maketitle

\begin{abstract}
Distributed system architectures such as cloud computing or the emergent architectures of the Internet Of Things, present significant challenges for security and privacy. Specifically, in a complex application there is a need to securely delegate access control mechanisms to one or more parties, who in turn can govern methods that enable multiple other parties to be authenticated in relation to the services that they wish to consume. We identify shortcomings in an existing proposal by Xu et al for multiparty authentication and evaluate a novel model from Al-Aqrabi et al that has been designed specifically for complex multiple security realm environments. The adoption of a Session Authority Cloud ensures that resources for authentication requests are scalable, whilst permitting the necessary architectural abstraction for myriad hardware IoT devices such as actuators and sensor networks, etc. In addition, the ability to ensure that session credentials are confirmed with the relevant resource principles means that the essential rigour for multiparty authentication is established.
\end{abstract}

\providecommand{\keywords}[1]
{
  \small	
  \textbf{\textit{Keywords---}} #1
}


%
\IEEEpeerreviewmaketitle
\section{Introduction}
As both individuals and organisations embrace the benefits of cloud computing infrastructure, more and more data storage and business process services are being transferred or established in clouds\cite{hill2013}. This shift from local to remote infrastructure drastically reduces the effort and expenditure\cite{bieter2010} required for system maintenance\cite{albeshri2010,calheiros2011}, enabling system users to concentrate on business concerns such as QoS and performance, etc.

The rapid uptake of services delivered via clouds has meant that matters of convenience have overtaken concerns about security and privacy. Distributed systems such as clouds inherently introduce security vulnerabilities that can be accessed remotely if insufficient security measures are in place. Cloud-based systems have specific limitations with regard to security and these are discussed by Al-Aqrabi et al\cite{alaqrabi2014a}.

There still remain enterprises and individuals who are reluctant to adopt cloud infrastructure, and the lack of awareness of secure methods of cloud adoption limits the business advantages that are available to users and enterprises\cite{sharma2018,chen2012}.

Choo\cite{choo2017} and Liu\cite{liu2015} both describe the need to ensure that both the providers and consumers of cloud computing have the appropriate mechanisms deployed for security and privacy. As such there has been considerable research activity\cite{thilakanathan,song2016,arya,celesti2010b} pertaining to the security of cloud applications and infrastructure\cite{ateniese2000,katz2003,rahulamathavan2014,yoon2005}.

\subsection{Multiparty service delivery and security}
As enterprises are beginning to become aware of the power of data collection, analysis, modelling and prediction, they are starting to realise systems that are a more faithful representation of business processes. This means that the underlying digital services must demonstrate both robustness and flexibility to tolerate new and unanticipated business scenarios. As such, the actual process flows may be difficult to predict in some instances, especially if a business offers bespoke services or products to customers, where a transaction may execute once only\cite{Georg2005}.

As a consequence of this, the eventual application that is delivered is underpinned by a collection of disparate services that are orchestrated at run-time, that may have origins in organisations that are heterogeneous.
Each of the host organisations will have adopted security measures that are unique to the enterprise, with the effect that an application composed of multiple services will thus present a number of different security realms.

Each realm typically consists of data that represents a collection of resource principals, that are registered with a trusted principal such as a certificate authority. The principals are governed by a set of security policies that control access to other services and resources within the scope of the application\cite{xu2012}. The certificate authority is deemed to be trustworthy across the application domain and is present to validate users and functions\cite{clercq2002}.

It is essential that each security realm is authenticated against to ensure that a principal has the appropriate security privileges to consume services marshalled by a security realm. The identity of a principal needs to be confirmed by the correct authentication process of the relevant realm so as to correctly identify and establish who the principal is. During the authentication process, security credentials that were given to the principal by the relevant security realm are used to authenticate it.

In the case of more complex application architectures, such as cloud-based services provision, each cloud may hide multiple instances of other clouds and/or services. It follows that not only will there be numerous authentication mechanisms to keep maintained, but they will have to be invoked dynamically at runtime on demand. If separate authentication processes are established  across disparate security realms, there is a potential for a significant increase in authentication workload and the consequential side-effects on network bandwidth and computational cycles.

The scenario where a multiparty session is composed of many two-party sessions is explored by Hada et al\cite{hada2002}, who demonstrates that there is a need for a protocol for multiparty session authentication. Thre is an inherent challenge here that it is not always possible for a session participant to establish whether another session participant is actually a member of the multiparty session in progress.

The rest of this article is organised as follows. First, we consider the main challenges for secure authentication in distributed systems infrastructure such as cloud computing, where multiple parties are present. In particular, we shall consider the key obstacles that are presented by environments that are composed of many different parties of varying capabilites such as with the Internet of Things (IoT).

Second, we shall briefly review some existing approaches to managing multiparty authentication, and critically discuss a model developed by Xu\cite{xu2012}. Third, we propose a distinct model for secure multiparty authentication that addresses shortcomings in current models, and explain how it can be deployed by way of an example. Finally, we describe some concluding remarks.
\section{Key challenges for multiparty environments}
Whilst multiparty authentication is a complex challenge in a multi-cloud environment, the complexity increases considerably when we consider the potential proliferation of devices in IoT systems. In such systems, there may be 1:1 mappings between system access devices and the clouds themselves, but there is also the additional potential complication of myriad hardware devices that possess varying degrees of functionality and capability. Such devices can be found in Wireless Sensor Networks for instance, which are often adaptive entities that can embrace the addition or removal of sensor nodes during operation.

If Gartner's prediction is true - ``Gartner, Inc. forecasts that 8.4 billion connected things will be in use worldwide in 2017, up 31 percent from 2016, and will reach 20.4 billion by 2020."\cite{gartner2017} - then the demand for authentication of devices will be a significant challenge to address, particularly since there will be insufficient capacity to manually authenticate even a fraction of the devices, and therefore some automation will be mandatory.

A fundamental challenge in a complicated environment such as the IoT or multi-clouds is the need to manage and assure the communications that enable the requisite authentication approvals to be enacted \cite{celesti2010b}.

The use of Single Sign On (SSO) has become a convention for users to conveniently access systems that are composed of multiple sub-systems, each of which may be a different application deployment. SSO removes the need for users to enter differing security credentials multiple times, and is enabled by the use of a key exchange mechanism to manage the provision of authentication credentials that have been certified by a named authority \cite{Sharma2016}, \cite{Schridde2010}. However, the relative simplicity of a mechanism to provide a secure method of key exchange is inadequate for the situation when we need multiple parties to be able to trust each other in a dynamic, heterogeneous environment, and therefore SSO is lacking in this regard.

In the next section we shall discuss existing approaches to multiparty authentication.
\section{Existing multi-party approaches}
%
In a multiparty concept, multiple parties can join or leave a session dynamically. The parties are allowed or removed from the session by a session authority. A simplified drawing showing the concept is presented in the below figure.

In this concept, a session authority controls the authentication of all session participants. Existing session participants can introduce new participants to the session authority for limited transactions. The session authority issues a secret session key to all running session participants.

In practice the session authority communicates with multiple session handlers. Whenever a new participant joins or existing one leaves, the key is refreshed and shared with the active session participants using forward security techniques \cite{alaqrabi2018}.

The session authority recognises the session participants with the help of participant IDs. A participant leaving the session cannot reuse its participant ID for re-entering.

Similarly, a reused ID will not be assigned to a new participant. The participants join through introduction only and need not share any secret artefacts to gain the session key. However, two participants acting as partners can share private keys using the Diffie-Hellman algorithm.


%
\begin{figure}[tb]
  \includegraphics[width=\linewidth]{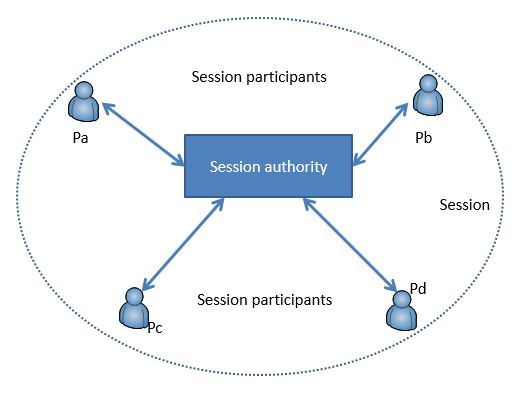}
  \caption{The multiparty session authentication concept.}
  \label{fig:maconcep}
\end{figure}


Prior work related to multiparty identity authentication in multi-cloud environments \cite{xu2012}, \cite{alaqrabi2018} has described models that utilise more sophisticated methods than password management through SSO. The concept of direct trust is introduced as part of their mechanism, with \cite{dai2011} introducing group credentials across security realms within inter-cloud communications.

The basic premise is that a new actor (either human or device) should be able to access a system as a result of a secret that is shared amongst a group. This is developed further by \cite{xu2012} and\cite{alaqrabi2018} whereby a third party, who is trusted, can discover and establish permissions on behalf of the actor.

In such a scenario, the system can observe that an actor is associated with a group that is trusted, and therefore will assign permissions based on the association. The use of trust relationships opens up the potential for actors to gain access to systems based upon their reputation, which is more akin to human social networks.

At any one time, it is conceivable that there will be many system access requests to complete in a secure and timely manner. This is one aspect of the emerging IoT scenarios that is particularly challenging, in terms of the sheer volume of potential requests that may exist. 

The Session Authority $(SA)$ is a role of fundamental importance in that it manages the confirmation and approval of access requests to the system. The $SA$ is preceded by a Multi-Party Session Handler ${MPSH}$ who formulates a queue of  requests for subsequent processing by the $(SA)$.

Since the workload of the $SA$ is likely to vary, and at least will be expected to scale upwards when an application grows by the addition of additional actors and their devices, one approach to deal with the elasticity in demand is to adopt an $SA$ Cloud $(SAC)$ as proposed by \cite{alaqrabi2012,alaqrabi2018} and illustrated in Figure \ref{fig:proposedmodel}. 
\section{A model for multiparty authentication}
\begin{figure}[tb]
  \includegraphics[width=\linewidth]{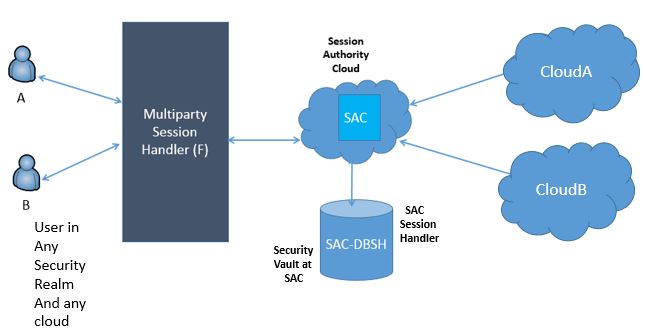}
  \caption{A multiparty session authentication model\cite{alaqrabi2018}.}
 \label{fig:proposedmodel}
\end{figure}

In this section, a model for dynamic authentication interactions in a distributed environment (in this example multiple clouds) is shown in Figure \ref{fig:proposedmodel}\cite{alaqrabi2018}. All members of multiple sub-domains may interact within a session and all such sessions are identified by the Session Authority Cloud ($SAC$). Session keys comprise of the root key of the cloud, a sub-domain key, and the portion identifying the session.

This means that there will be multiple session keys valid for a session, each having a common field for the session, but varying fields for cloud root keys and sub-domain keys. There is no need for any negotiation among the clouds because the session authority cloud is aware of all the clouds and their sub-domains.

The schematic of the robust multiparty model is shown in Figure \ref{fig:proposedmodel}.
The ground rules of the proposal are:
\begin{itemize}
\item Each session participant should be a tenant of at least one cloud in the multi-cloud framework controlled by the session authority cloud.
\item If a potential participant is not a cloud member, the introducing participant will have to share credentials with it for joining its own cloud.
\item Each session will have multiple valid keys. While the session key field will be common (refreshed on change of no of participants), the cloud root keys and sub-domain (security realm) keys will vary depending upon the membership profiles of the participants.
\end{itemize}
The session begins with a user having membership in any security realm (cloud $C$) that the trusted principal recognises. We assume to provide access to some resources such as database objects in clouds $A$ and $B$ if the $SAC$ approves the request forwarded by the principal.

It is also assumed that $SAC$ will not entertain any request not forwarded by the principal. The user requesting access is neither a member of cloud $A$ nor a member of cloud $B$. In essence, the user is a member of a security realm that is a different cloud (cloud $C$), which is trusted by the $SAC$ (which means that the third cloud is a member of the $SAC-DB$).

Most importantly, the principal should recognise who is the user because the $SAC$ trusts the principal for accepting the session request. Hence, the only way the user can gain access to database files on clouds $A$ and $B$ is to send a request to the session authority cloud through the session handler $F$.

The session handler will only forward requests of the trusted principal and hence the requests need to be forwarded through the login of the trusted principal.

The principal $A$ places a request to the session handler to gain access to resources $A$ and $B$ for the user (the user knows their URLs but does not have any access to them).

On the request of the session handler ($F$), the principal $A$ shares the root and sub domain keys of the user. These keys may be viewed as two packets of a finite size (for example, 1024 Bytes each). 

The $SAC$ checks the keys with the help of a database $SAC-DB$ to assist it (may be viewed as a huge security vault having all root and sub-domain keys of the clouds registered with it).

On confirmation from $SAC-DB$, the $SAC$ approves access to database files $A$ and $B$ stored on clouds $A$ and $B$ respectively. It forwards its approval to the $SAC$’s session handler $(SAC-SH)$. The $SAC-SH$ may be viewed as a separate dynamic database that caches all approvals from the $SAC$ and forwards them to respective clouds for opening the accesses.

Cloud $A$ stores the session ID and key in its registry or cache and then sends a response to $SAC$. $SAC$ sends a reply for session approval to $F$. Then, $F$ sends a response for session approval to access the application on cloud $A$ or $B$.

\begin{figure}[tb]
  \includegraphics[width=\linewidth]{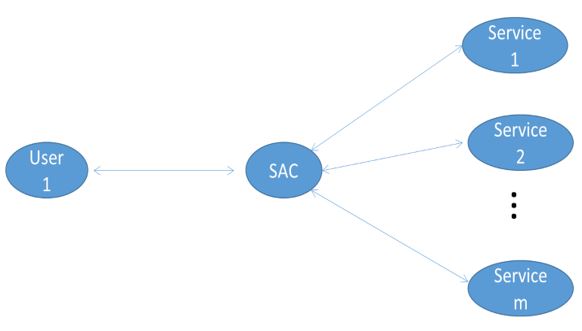}
  \caption{A worst case scenario where either $n$ users and one service, or there are 2 users and 2 services.}
 \label{fig:worstca}
\end{figure}

\begin{figure}[t]
  \includegraphics[width=\linewidth]{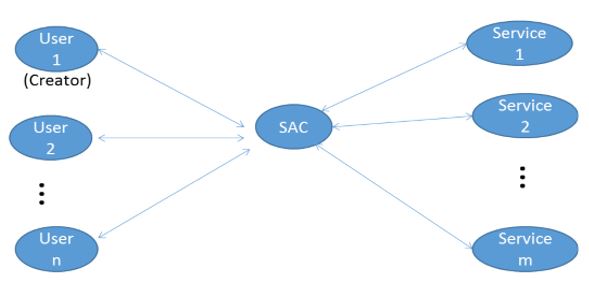}
  \caption{A best case scenario where there are $n$ session users and $m$ services, where both $n$ and $m$ are much greater than 2.}
 \label{fig:bestca}
\end{figure}

Typically, the authentication processes occur between $SAC$ and session members (users and services). Figure \ref{fig:worstca} shows a worst case scenario, where there is only one user and $m$ services in a session. 
Conversely, worst case scenarios exist when there are either $n$ users and one service, or there are 2 users and 2 services. In each of these cases, the authentication process cannot be simplified and the $SAC$ process offers no significant advantage over direct authentication. In addition, the two-party session technique does not address the issue of different Cross-Realm Authentication, which requires credential conversion and the establishment of authentication paths.
On the other hand, Figure \ref{fig:bestca} shows a best case scenario for the multiparty model. In this case there are $n$ session users and $m$ services, where both $n$ and $m$ are much greater than 2. The benefit is obtained since each user will be able to access all of the $m$ services. 
Without this model, the authentication processes must be performed within sessions because each user has to be authorised by all of the services that they want to access. 

\section{Different security realms}
If we consider the situation whereby authentication is required across two different security realms, it is necessary to a) convert credentials so that they can be shared, and b) define specific authentication paths between the relevant realms. Existing methods for two-party session authentication are not able to resolve such scenarios.

One approach is to use federated authentication, although the negotiation amongst parties is time intensive, especially when one or more parties requires the authentication path to be amended or augmented to their satisfaction. The complexity of this situation escalates rapidly as more parties are added, either as actors wanting access, or myriad services distributed across differing security realms.

The fact that existing approaches do not comprehensively address the increasingly prevalent situation where multiple actors have a need to access multiple services, across multiple security realms (such as a multi-cloud environment), means that business models that are based upon the distributed provision of services are potentially hampered by the ability to scale authentication at a less than optimal rate. We foresee this as a significant barrier for secure IoT, as the orchestration of services (latterly microservice architectures\cite{shadija2017,hill2015}) held within containers is a natural progression as interest develops for scalable business models and infrastructure.
\section{Evaluating multiparty models}
%
We now proceed to examine the multiparty scenario described by Xu\cite{xu2012} in order to understand some of the pertinent challenges faced by multiparty authentication models.
The multiparty model of Xu\cite{xu2012} employs a $SA$ entity that oversees the authorisation of sessions against session requests from existing partners. 
With this arrangement a session can be established by an actor that is able to provide an instance ID; this means that there is no need to identify the resources to be accessed, nor to initiate contact with their principals.

If two session parties, $A$ and $B$ have already communicated and established trust, either $A$ or $B$ can introduce a new party, $C$. The $SA$ will approve the request merely because $A$ and $B$ are in a current session, irrespective of the means used to access the resources, which may be either legitimate or unauthorised.

A secret session key is provided to $C$ to join the session. If any one of the parties leave the session (for instance $A$), the secret key will be refreshed such that $A$ cannot rejoin on its own (silently). If $A$ does attempt to join, it will be detected and considered a potential security breach.

Thus, $A$ needs to seek sponsorship of $B$ or $C$ as they are already authenticated in a session, even though this $C$ was introduced to the session by $A$ earlier.
Therefore, the model proposed by Xu enables session participants to introduce new parties, a decision that is endorsed by the $SA$ without any conditions.

Furthermore, the $SA$ has no concern what each party is accessing within a session. This is significant in the following situation. If parties $A$, $B$, and $C$ are accessing a restricted database, the $SA$ does not seek approval from the owner of the database.

Hence, if $C$ becomes the provider of a session instance ID, as part of an inside attack upon the database, the $SA$ will honour the request as no further checks are made. In fact, the database owner is an invisible security realm for the $SA$, as it is only concerned with a session $S$ via the session ID and instance ID.

If all three parties were adversarial attackers and were successful at breaching the database be creating a $SI$ and a $CI$, any request for part $D$ to join the session would be approved by the $SA$ and the actual authority of the database would never be contacted to confirm or deny credentials.

To summarise the observations so far with respect to the multiparty model of Xu\cite{xu2012}:
\begin{enumerate}[(a)]
\item The $SA$ has no knowledge about resources used in the session and is only concerned with $CI$ and $SI$;
\item The $SA$ has no access to knowledge of the principal owners (the actual authority) of resources for a session;
\item The $SA$ never contacts the principal resource owners as long as an instance key is provided by a party that already has access to the resource; 
\item The $SA$ has no other mediator to assist with authentication checks;
\item The $SA$ cannot control the addition of access to resources during an executing session.
\end{enumerate}

\begin{figure}[t]
  \includegraphics[width=\linewidth]{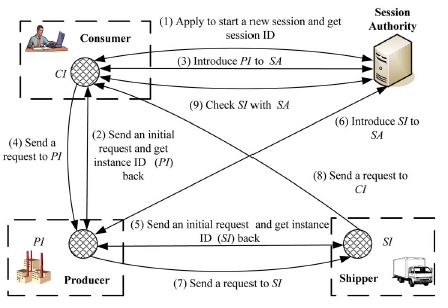}
  \caption{Multiparty business scenario as described by Xu\cite{xu2012}.}
\end{figure}
\begin{figure}[t]
  \includegraphics[width=0.96\linewidth]{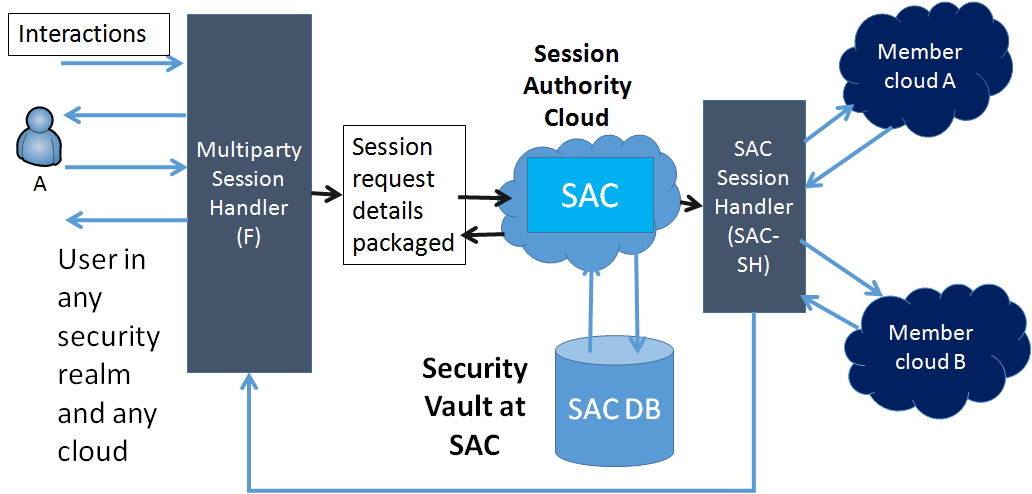}
  \caption{Multiparty authentication model as described by Al-Aqrabi\cite{alaqrabi2018}.}
 \label{fig:alaqrabi}
\end{figure}

In contrast, we shall now consider the multiparty model that employs a Session Authority Cloud ($SAC$) entity\cite{alaqrabi2018}, illustrated in Figure \ref{fig:alaqrabi}.

The $SAC$ differs from the $SA$ of Xu\cite{xu2012} in that it includes a security vault database ($SAC-DB$) to facilitate authentication checks beyond that of session coordination only. This database provides knowledge of what activities are being performed within a session, which removes the potential vulnerability of the $SA$ having the power to grant session access to a proposer without requiring any resource checks.

The $SAC-DB$, together with the $SAC Session Handler$ ($SAC-SH$), ensure that session participants are authenticated only when all details of the required resources are confirmed, rather than relying only on the collection of session keys as the sole authentication mechanism. If the resource checks fail, the $SAC$ will not allow the session to execute. As such, the $SAC$ contacts the destination resources for confirmation, which is absent from the previous model\cite{xu2012}.

The fact that the $SA$ will honour requests for access without further authentication, means that Xu's model cannot protect an application from multiparty adversarial attacks, which are typically initiated from the inside. As the resource owner belongs to a realm that the $SA$ is not concerned with (as it requires only a session and instance ID), one nefarious party can grant access to another.

This could be an inside agent that enables an additional external agent to compromise the application for malicious purposes. We feel that this is a major barrier to the adoption of such a model, particularly for IoT architectures where there is a reasonable expectation that additional devices will be added to the system in the future, that will require authentication through verified credentials.

Approval is requested by the $SAC$ from the resource owner (e.g., database owner), who is a security realm principal and external to the session. Furthermore, the $SAC$ takes individual approvals for the resources accessed and hence, a new resource access request cannot be added during a session that is executing.

For example, if we consider the playing of video content from embedded links within a database: as soon as an embedded video link is launched, an approval screen with a pop-up directing the users to involve the $SAC$ again, who in turn will seek its access from the principal owner of the videos. The $SAC$ is supported by the $SAC-DB$ and $SAC-SH$ to ensure that the principals of the resources are contacted for appropriate authentication and approval.

Alternatively, an IoT based network may utilise wireless communication to enable wireless sensors and other mobile devices to move into and out of the scope of an application. This might enable a network to utilise opportunistic sensing, or packet transport from IoT appliances such as vehicles. In this situation, it will be necessary to establish a trust relationship that may be transient and limited in its ability to access certain resources within the application.
In this case, the $SAC$ again is assisted by the $SAC-DB$ and $SAC-SH$, so that access is provided to the requisite parties in a secure manner.

\section {Contribution}
%
A more substantial process of establishing a multiparty session is a key differentiator between the two models. This extra emphasis within the second model\cite{alaqrabi2012} ensures that the trust relationship is more secure from the outset.

Access to resources is strictly limited to the approvals by the resource principals of  cloud $A$ and cloud $B$, who are named ownership entities rather than an adversary who succeeds in generating an instance ID. In addition, even if the principals are not in the session, they can monitor it from outside.
        

Finally, we consider the more demanding scenario of large collections of IoT devices being present as part of emergent system architectures, whose fundamental capabilities have to be based upon secure scalability. The $SA$ entity of Xu's model serves as a constraint in that is cannot function without intervention to generate and share instance IDs.

Thus, the multiparty model proposed in\cite{alaqrabi2012} presents two siginificant contributions as follows:
\begin{enumerate}
\item The employment of a $SAC$, together with a $SAC-DB$ and $SAC-SH$ ensures that session authority is granted with full cooperation of the resource principals, which provides more rigour when establishing a trust relationship at the outset;
\item Scalability of authentication requests is catered for through the abstraction of the $SAC$; hardware devices such as actuators, sensor networks, etc., can be fully hidden behind the $SAC$ and the $SAC-SH$ entities.
\end{enumerate}
Together, contributions (1) and (2) provide a more robust base upon which to consider the security of multiparty application architecture.

\section {Conclusion}
In conclusion this article examines approaches to multiparty authentication through the lens of applications that have a future requirement to scale upwards, such as the emergent growth of IoT networks. Such architectures are dynamic, must exhibit elasticity, and also present a multitude of potential security vulnerabilities. Nonetheless, the appeal of distributed hardware that is networked provides a compelling motivation for business and individual users alike to adopt such technologies.

We have selected two proposed models of multiparty authentication and examined their suitability with respect to the IoT scenario described above. We have identified that models which can be utilised for Service Oriented Architectures need to ensure that the authentication controls, policies and protocols need to protect against the inevitable multiparty interactions that will have to be satisfied.

We examine a novel multiparty authentication approach\cite{alaqrabi2012,alaqrabi2014} that provides a robust mechanism for marshalling security interactions with distributed infrastructures such as multi-cloud and IoT networks. This model is pertinent when multiple members of heterogeneous security realms have a desire to access a variety of services, under the governance of a trusted principal. The model thus enables authentication dynamically during execution of a multiparty application, whilst ensuring that there is a minimal requirement to convert security credentials as the services are accessed.
A major motivation for the adoption of this model is that of authentication simplification between two or more services that are unrelated, before the services are permitted to exchange data. Only when authentication checks have been made with resource principals is data exchange permitted.

We intend to develop this work to identify and address new challenges posed by heterogeneous IoT environments. Key areas of focus are as follows:
\begin{itemize}
\item Developing proofs of secure authentication protocols for multiparty scenarios;
\item Using hardware-in-the-loop to deploy IoT networks, enabling authentication overheads to be evaluated for a variety of use cases, particularly in relation to different networking protocols such as LoraWAN and Zigbee;
\item Monitoring and assessing the impact of energy consumption of the authentication protocols upon constrained hardware resources;
\item Investigating methods for the visualisation of authentication and multiparty behavioural signatures to improve resilience towards more sophisticated adversarial attacks;
\item Developing mechanisms to automate trust formation through authentication for resource-constrained hardware.
\item Exploring the use of SDN to manage network traffic across IoT devices in a multiparty authentication scheme\cite{baker2012}.
\item Developing resource scheduling algorithms to manage multiparty authentication workload across multiple clouds and IoT hardware\cite{sotiriadis2013}.
\end{itemize}

\end{document}